\documentclass[aps,prb,twocolumn,superscriptaddress,showpacs]{revtex4}
\usepackage{graphicx}
\usepackage{latexsym}
\usepackage{amssymb}
\usepackage{amsmath}
\usepackage{amsfonts}
\usepackage{bm}
\usepackage{multirow}
\usepackage{color}
\usepackage{comment}
\newcommand{\ii}{\mathrm{i}}

\newcommand{\SO}{\mathrm{SO}}

\newcommand{\SU}{\mathrm{SU}}
\newcommand{\U}{\mathrm{U}}
\newcommand{\Sp}{\mathrm{Sp}}

\newcommand{\MF}{\mathrm{MF}}
\newcommand{\FM}{\mathrm{FM}}
\newcommand{\tr}{\mathrm{tr}}

\newcommand{\beq}{\begin{equation}}
\newcommand{\eeq}{\end{equation}}
\newcommand{\beqn}{\begin{eqnarray}}
\newcommand{\eeqn}{\end{eqnarray}}

\DeclareMathAlphabet{\mathbbold}{U}{bbold}{m}{n}

\def\SU{{\rm SU}}

\def\U{{\rm U}}

\begin{document}

\title{Ferromagnetism and Spin-Valley liquid states in Moir\'{e} Correlated Insulators}


\author{Xiao-Chuan Wu}
\affiliation{Department of Physics, University of California,
Santa Barbara, CA 93106, USA}

\author{Anna Keselman}

\affiliation{Kavli Institute of Theoretical Physics, Santa
Barbara, CA 93106, USA}

\author{Chao-Ming Jian}
\affiliation{Kavli Institute of Theoretical Physics, Santa
Barbara, CA 93106, USA}

\author{Kelly Ann Pawlak}
\affiliation{Department of Physics, University of California,
Santa Barbara, CA 93106, USA}

\author{Cenke Xu}
\affiliation{Department of Physics, University of California,
Santa Barbara, CA 93106, USA}

\begin{abstract}

Motivated by the recent observation of evidences of ferromagnetism
in correlated insulating states in systems with Moir\'{e}
superlattices, we study a two-orbital quantum antiferromagnetic
model on the triangular lattice, where the two orbitals physically
correspond to the two valleys of the original graphene sheet. For
simplicity this model has a $\SU(2)^s \otimes \SU(2)^v$ symmetry,
where the two $\SU(2)$ symmetries correspond to the rotation
within the spin and valley space respectively. Through analytical
argument, Schwinger boson analysis and also DMRG simulation, we
find that even though all the couplings in the Hamiltonian are
antiferromagnetic, there is still a region in the phase diagram
with fully polarized ferromagnetic order. We argue that a Zeeman
field can drive a metal-insulator transition in our picture, as
was observed experimentally. We also construct spin liquids and
topological ordered phases at various limits of this model. Then
after doping this model with extra charge carriers, the system
most likely becomes spin-triplet/valley-singlet $d+id$ topological
superconductor as was predicted previously.

\end{abstract}

\maketitle

\section{Introduction}

Recently a series of surprising correlated physics such as
superconductivity and insulators at commensurate fractional charge
fillings have been discovered in multiple systems with Moir\'{e}
superlattices~\cite{wangmoire,mag01,mag02,young2018,TLGSC,FM,kimtalk,TDBG1,TDBG2}.
These discoveries have motivated very active theoretical
studies~\cite{xuleon,senthil,kivelson,fu,vafek,phillips,phillips2,baskaran,yang,louk,fu2,fu3,ashvinyou,martin,zhang,bernevig,scalet,senthil2,senthil3,vafek2,fczhang,subir,balents2,dai,xuwire,bruno,zaletel,TDBGt,dai2,TDBGt2}.
For different reasons, these systems can have narrow electron
bandwidth near charge
neutrality~\cite{flat1,flat2,flat3,flat4,TDBG1,TDBG2}, hence
interaction effects are significantly enhanced.
A consensus of the mechanism for the observed insulator and
superconductor has not yet been reached. A minimal two-orbital
extended Hubbard model on the triangular lattice was proposed in
Ref.~\onlinecite{xuleon} (physically the orbital space corresponds
to the space of two Dirac valleys in the original Brillouin zone
of graphene), which at least describes the trilayer graphene and
hexagonal BN heterostructure (TLG/h-BN)~\cite{wangmoire,TLGSC}, as
well as the twisted double bilayer graphene
(TDBG)~\cite{kimtalk,TDBG1,TDBG2} with certain twisted angle and
out-of-plane electric field (displacement field), since in these
cases there is no symmetry protected band touching below the fermi
energy, and the isolated narrow band has trivial quantum valley
topological number~\cite{senthil,band3,band4,band5,TDBGt,dai2}.
This minimal model would then naturally predict either a
spin-triplet~\cite{xuleon} or spin-singlet~\cite{kivelson} $d+id$
topological superconductor, depending on the sign of the on-site
Hund's coupling.

Evidence of spin-triplet pairing predicted
previously~\cite{xuleon} was recently found in
TDBG~\cite{kimtalk,TDBG2}. Evidence of ferromagnetic correlated
insulator at half filling away from charge neutrality was
discovered in the same system~\cite{kimtalk,TDBG1,TDBG2}. In TDBG,
besides clear ferromagnetism signature observed at the
$1/2$-filling insulator~\cite{kimtalk,TDBG1,TDBG2}, it was also
observed that correlated insulators at $1/4$ and $3/4$ fillings
emerge under inplane magnetic field~\cite{TDBG2}, whose main
effect is likely just a spin-polarizing Zeeman effect. This
observation implies that the TDBG at $1/4$ and $3/4$ filling is
rather close to a ferromangetic correlated insulator, and a Zeeman
field would drive a metal-insulator transition. Evidence of
ferromagnetism at the insulating phase with 3/4 filling away from
the charge neutrality was also found in another Moir\'{e}
system~\cite{FM}.

Motivated by these experiments, in this work we investigate a
quantum spin-valley model on the triangular lattice with one
fermion per site, which corresponds to either $1/4$ filling or
$3/4$ filling on the Moir\'{e} superlattice. The Hamiltonian of
this model reads \beqn H = \sum_{<i,j>} \sum_{a,b = 1}^3 J
T^{ab}_{i} T^{ab}_j + J^s \sigma^a_{i} \sigma^a_{j} + J^v \tau^b_i
\tau^b_j, \label{H}\eeqn where $\sigma^a$ and $\tau^b$ are Pauli
operators in the spin and valley spaces, and $T^{ab} = \sigma^a
\otimes \tau^b$. When $J^s = J^v = J$, this model becomes the
$\SU(4)$ quantum antiferromagnetic model with fundamental
representation on each site. The $\SU(4)$ symmetry is broken by
the Hund's coupling~\cite{xuleon}, which in general makes $J^v > J
> J^s$, if we choose the standard sign of the Hund's coupling which
favors large spin on each site. But we assume that the $\SU(4)$
breaking effect is not strong enough to change the sign of $J^s$,
$J^v$ and $J$, namely we keep all three coupling constants
positive, $i.e.$ antiferromagnetic. Indeed, since the Hund's
coupling originates from the exchange coupling which involves
overlap between wave functions at the two valleys, the Hund's
coupling should be a relatively weak effect since the inter-valley
wave function overlap is expected to be small because large
momentum transfer between the two valleys is suppressed by the
long wavelength modulation of the Moir\'{e} superlattice. For
simplicity we ignore other mechanisms that break the $\SU(4)$
symmetry, such as valley-dependent hopping~\cite{senthil}, hence
in the spin-valley model Eq.~\ref{H} the valley space has its own
$\SU(2)^v$ symmetry.

\section{Derivation of the Spin-Valley model}

The model Eq.~\ref{H} can be derived in the standard perturbation
theory starting with a Hubbard model plus an on-site Hund's
coupling on the triangular lattice. As we mentioned in the
introduction, this model at least applies to
TLG/h-BN~\cite{wangmoire,TLGSC} and
TDBG~\cite{kimtalk,TDBG1,TDBG2} with certain twisted angle and
displacement field, since in these cases there is no symmetry
protected band touching below the fermi energy, and the narrow
band with correlated physics has trivial quantum valley
topological number (although the exact valley topological number
computed depends on the models used in the
literature)~\cite{senthil,band3,band4,band5,TDBGt,dai2}:
\begin{flalign}
H&=H_{t}+H_{U}+H_{V}+\ldots \\ H_{t}&=-t\sum_{\left\langle i,j
\right \rangle
}\sum_{\alpha=1}^{4}\left(c_{i,\alpha}^{\dagger}c_{j,\alpha}+\textrm{H.c.}\right),\\H_{U}&= U \sum_{j}\left(n_{j}-1\right)^{2},\\
H_{V}&=-V\sum_{j}\left( \vec{\hat{\sigma}}_j \right)^2 +
V\sum_{j}\left( \vec{\hat{\tau}}_j \right)^2 \label{hund},
\end{flalign}
where $c_{j,\alpha}^{\dagger},c_{j,\alpha}$ are electron creation
annihilation operators which have four flavors including both the
spin and valley indices,
$\hat{n}_{j}=\sum_{\alpha}c_{j,\alpha}^{\dagger}c_{j,\alpha}$ is
the total particle number per site, and
$\frac{1}{2}\hat{\sigma}_j^{a}=
\frac{1}{2}c_{j}^{\dagger}\sigma^{a}c_{j}$ is the total on-site
spin operator. The $H_{U}$ term is the on-site repulsive
interaction, and we assume the most natural sign of the Hund's
coupling with $V
> 0$, as a result of exchange interaction. We treat the kinetic
term $H_t$ perturbatively. This amounts to integrating out the
charge degree of freedom to obtain an effective spin-valley model
in the correlated insulator phase.

We follow the standard approach of degenerate perturbation theory.
At quarter-filling, the ground state of $H_{U}+H_{V}$ has
precisely one electron per site, and the projection operator to
the ground state manifold reads \beqn \mathcal{P} = \prod_{j}
(-1)\frac{1}{6}n_{j}\left(n_{j}-2\right)\left(n_{j}-3\right)\left(n_{j}-4\right).
\eeqn Considering any pair of nearest neighbor sites on the
Moir\'{e} superlattice, the ground state manifold can be further
divided into four sectors which correspond to spin-singlet/triplet
and valley-singlet/triplet states. We can write \beqn \mathcal{P}
= \mathcal{P}_{ss} +\mathcal{P}_{st} +\mathcal{P}_{ts}
+\mathcal{P}_{tt}, \eeqn where (for example) $\mathcal{P}_{st}$
means the projection to spin-single/valley-triplet states. The
effective Hamiltonian can be calculated as \beqn
H_{\textrm{eff}}=\mathcal{P}H_{t}\frac{1}{E_{0}-H_{U}-H_{V}}H_{t}\mathcal{P},
\eeqn where $E_{0}$ is the ground state energy for the two-site
problem. A detailed analysis of the intermediate states
considering the virtual hopping process can be found in
Ref.~\onlinecite{xuleon}. We find that only $\mathcal{P}_{st}$ and
$\mathcal{P}_{ts}$ contribute to the effective Hamiltonian which
takes a diagonal form in this basis \beqn
H_{\textrm{eff}}=-\frac{2t^{2}}{U+4V}\mathcal{P}_{st}-\frac{2t^{2}}{U-4V}\mathcal{P}_{ts}.
\eeqn Rewritten in terms of the $\SU(4)$ generators on the nearest
neighbor sites, the effective Hamiltonian is equivalent to the
spin-valley model Eq.~\ref{H} with the coupling constants given by
 \beqn
 J^{s} &=& 
 J - \frac{t^{2}}{U}\left( \frac{2 V}{U} + O \left(\frac{V}{U} \right)^2 \right),\cr\cr
 J^{v} &=& 
 J + \frac{t^{2}}{U}\left( \frac{2V}{U} + O \left(\frac{V}{U} \right)^2
 \right), \cr\cr
 J &=& 
 \frac{t^{2}}{4U} \left( 1 + O \left(\frac{V}{U} \right)^2
 \right).
 \eeqn There is a $Z_2$ symmetry
regarding the sign of the Hund's coupling. The coupling constants
transform as $J^{s}/J\leftrightarrow J^{v}/J$ when we change
$V\leftrightarrow-V$, as is naturally expected from the form of
the Hund's coupling Eq.~\ref{hund}.

\section{The $\FM \otimes 120^\circ$ state}

At least in certain limit, $i.e.$ $J^v \gg J \gg J^s > 0$, it is
fairly easy to see why ferromagnetism would emerge in model
Eq.~\ref{H} with all antiferromagnetic coupling constants. First
of all, the following state will always be an eigenstate of the
Hamlitonian: \beqn |\Psi_{\mathrm{FM}}\rangle = \left( \prod_i |
\sigma^z_i = +1 \rangle \right) \otimes | \mathrm{AF} \
\mathrm{of} \ \vec{\tau}_{\{ i \}} \ \rangle. \label{wf}\eeqn This
state is a direct product of two parts: the first part is a
fully-polarized ferromagnetic state of the spin $\vec{\sigma}_i$
space; the second part is the ground state of the nearest-neighbor
antiferromagnetic quantum Heisenberg model on the triangular
lattice in the $\vec{\tau}_i$ space. Although we cannot write down
the explicit form of the exact microscopic wave-function $|
\mathrm{AF} \ \mathrm{of} \ \vec{\tau}_{\{ i \}} \ \rangle$, we do
know that this state has a $120^\circ$ antiferromagnetic order
with reduced moment due to quantum fluctuation and geometric
frustration. This state Eq.~\ref{wf} is always the eigenstate of
Eq.~\ref{H} because a fully polarized ferromagnetic spin state is
the eigenstate of operator $\vec{\sigma}_i \cdot \vec{\sigma}_j$
on every link $<i,j>$. Then in the limit of $J^v \gg J \gg J^s >
0$, this eigenstate $|\Psi_{\mathrm{FM}}\rangle$ is also the
ground state, because intuitively on every link the spin
$\vec{\sigma}_i$ will see a background ``effective" ferromagnetic
coupling \beqn J_{\mathrm{eff}} = J^s + J \langle \vec{\tau}_i
\cdot \vec{\tau}_j \rangle. \eeqn Because $\langle \vec{\tau}_i
\cdot \vec{\tau}_j \rangle < 0$ for the $120^\circ$ state of
$\vec{\tau}_i$, for large enough $J$ the spins will see an
effective ferromagnetic coupling, even though in the original
model Eq.~\ref{H} all the couplings are antiferromagnetic.

With a fixed large $J^v$, while increasing $J^s$, eventually $J
\langle \vec{\tau}_i \cdot \vec{\tau}_j \rangle$ will not be
strong enough to overcome the antiferromagnetic coupling $J_s$,
hence we expect to see a transition from the ``$\FM \otimes
120^\circ$" state to another state without ferromagnetic order.
Numerically~\cite{THA01}, $\langle \vec{\tau}_i \cdot \vec{\tau}_j
\rangle$ is found to be $\sim -0.73$ for the triangular lattice
quantum antiferromagnet. If we evaluate the energy of
$|\Psi_{\mathrm{FM}}\rangle$ in Eq.~\ref{wf}, while increasing
$J_s/J$, this state is no longer the ground state when $J_s/J
> 0.73$. Hence the intuitive argument gives an upper bound for
the transition point of $J^s/J$.

{\it ---DMRG simulation of the spin-valley model}


\begin{figure}
\includegraphics[width=\linewidth]{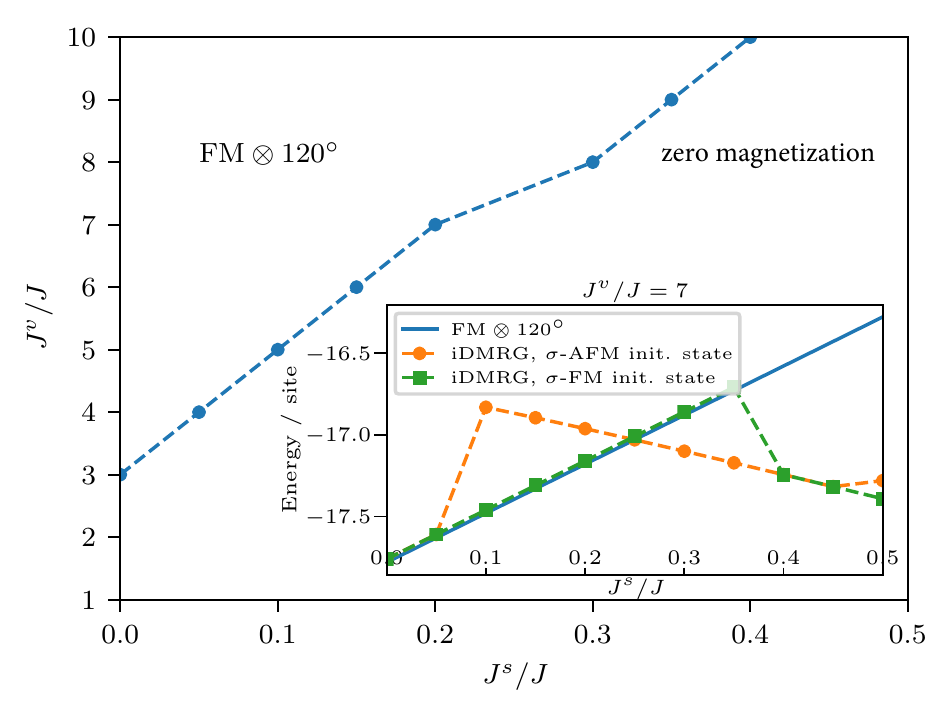}
\caption{Phase boundary of the $\FM \otimes 120^\circ$ state
obtained using DMRG on infinite cylinders with circumference
$L_y=6$. The inset shows the energy per site obtained for
$J^v/J=7$ as function of $J^s/J$. Green squares (orange circles)
indicate the energies obtained using iDMRG when the spins are
initialized in a FM (AFM) product state. The solid blue line
indicates the energy expected for the $\rm{FM}\otimes120^{\circ}$
state. } \label{pd_fig}
\end{figure}

We now provide numerical evidence for the fact that the $\FM
\otimes 120^\circ$ state is the ground state in the $J^v \gg J \gg
J^s$ limit and obtain the phase boundary of this state. To this
end we use the density matrix renormalization group (DMRG)
method~\cite{white1992,schollwoeck2005}. We note that in finite
systems, boundaries can introduce strong oscillations in the
expectation value of $\langle \vec{\tau}_i \cdot \vec{\tau}_j
\rangle$ on nearest-neighbor bonds (which are expected to be
uniform when the valley degree of freedom is in the $120^{\circ}$
state), and thus affect the effective coupling seen by the spins.
To avoid such boundary effects we use infinite
DMRG~\cite{McCulloch2008}. To observe uniform bond expectation
values when the valley degree of freedom is in the $120^{\circ}$
state wide enough cylinders have to be considered. We perform our
analysis on cylinders of circumference $L_y=6$, for which we
obtain mean bond expectation value $\langle \vec{\tau}_i \cdot
\vec{\tau}_j \rangle\approx-0.74$, consistent with
Ref.~\onlinecite{THA01}, with spatial variations below one
percent.

For our numerical simulations we use the ITensor
library~\cite{ITensor}. We assume a 3-site unit cell along the
cylinder to allow for the formation of a $120^{\circ}$ state in
the valley and/or spin degrees of freedom. The valley degree of
freedom is initialized in the total $\tau^z=0$ sector and $\tau^z$
quantum number conservation is used. The spin degree of freedom is
initialized either in the fully polarized state, or a classical
anti-ferromagnetic state with total $\sigma^z=0$. The maximal bond
dimension in our simulations is $M=1000$.

We find that at large $J^v$ and small $J^s$, the system indeed
converges to a {\it fully polarized} spin-FM and a
$120^{\circ}$-valley ordered state, independent of the initial
conditions. At larger $J^s$ we observe a state {\it without any
net magnetization}. At this stage we cannot conclude as to the
nature of the entire region with zero magnetization (with larger
$J^s$ and $J^s < J$, for the total $\sigma^z=0$ sector and going
to bond dimensions of up to 4000 we have not identified a clear
order for the spin degree of freedom), but in the next section we
will propose some possible interesting liquid states and
topological orders for this region of phase diagram. The ground
state energy obtained using iDMRG for a fixed $J^v/J=7$ as
function of $J^s$, for the two initial states, is shown in the
inset of Fig.~\ref{pd_fig}. The solid blue line on the same plot
indicates the energy expected for the $\rm{FM}\otimes120^{\circ}$
state, calculated using a uniform bond expectation value $\langle
\vec{\tau}_i \cdot \vec{\tau}_j \rangle \approx-0.74$ that we
obtain for the $120^{\circ}$ state on infinite cylinders of
circumference $L_y=6$ as mentioned above. We estimate the position
of the phase boundary for each $J^v/J$ to be at the $J^s/J$ for
which the lowest energy obtained using iDMRG drops below the one
expected for the $\rm{FM}\otimes120^{\circ}$ state. The resulting
phase boundary as function of $J^v/J$ and $J^s/J$ is shown in the
main Fig.~\ref{pd_fig}. Our results for the ground state energy
and the magnetization across the phase boundary both suggest that
the transition between the ferromagnetic order and the paramagnet
is a first order level-crossing.

{\it ---Schwinger boson analysis}

We can also construct the $\FM \otimes 120^\circ$ state using the
Schwinger boson formalism. We first define a four component
Schwinger boson $b_{j,\alpha}$ on every site which forms
fundamental representation under both the spin and valley $\SU(2)$
symmetry, and also a fundamental representation of the enlarged
$\SU(4)$ symmetry. The Schwinger boson Hilbert space is subject to
a local constraint \beqn \sum_{\alpha = 1}^4
b^\dagger_{j,\alpha}b_{j,\alpha} = \kappa. \label{cons}\eeqn
Physically $\kappa=1$, but in the Schwinger boson mean field
calculation $\kappa$ is often treated as a tuning parameter. The
Hamiltonian Eq.~\ref{H} can be reorganized into the following
form: \beqn H &=& J_{ts} \left( \vec{\Delta}^{ts \dagger}_{ij}
\cdot \vec{\Delta}^{ts}_{ij} \right) + J_{ss} \left( \Delta^{ss
\dagger}_{ij} \Delta^{ss}_{ij} \right) \cr\cr &+& J_{st} \left(
\vec{\Delta}^{st \dagger}_{ij} \cdot \vec{\Delta}^{st}_{ij}
\right) + J_{tt} \tr \left( \Delta^{tt \dagger}_{ij} \cdot
\Delta^{tt}_{ij} \right). \label{H1} \eeqn The operator
$\vec{\Delta}^{ts}_{ij}$ and $\vec{\Delta}^{st}_{ij}$ are the
spin-triplet/valley-singlet, and spin-singlet/valley-triplet
pairing operator between Schwinger boson $b_\alpha$ on site $i,j$:
\beqn \left( \vec{\Delta}^{ts}_{ij}, \vec{\Delta}^{st}_{ij}
\right) = b^t_i \left( \ii \sigma^{32}, \sigma^{02}, \ii
\sigma^{12}, \sigma^{23}, \ii \sigma^{20}, \sigma^{21} \right)
b_j, \eeqn where $\sigma^{ab} = \sigma^a \otimes \tau^b$,
$\sigma^0 = \tau^0 = \mathbf{1}_{2\times 2}$. Then $\Delta^{ss}$
and $\Delta^{tt}$ are the singlet/singlet, and triplet/triplet
pairing respectively, for example $\Delta^{ss}_{ij} = b^t_i
\sigma^{22} b_j$.

In Eq.~\ref{H1}, \beqn && J_{ts} = - \frac{1}{4}(3J_v + 3J - J_s),
\ \ J_{st} = - \frac{1}{4} (3 J +3 J_s - J_v), \cr\cr && J_{ss} =
\frac{3}{4} (3J - J_s - J_v), \ \ J_{tt} = \frac{1}{4}(J + J_s +
J_v). \eeqn With the most natural parameter region $J_v > J > J_s
> 0$, $J_{ts}$ is always negative, and it corresponds to the
strongest mean field channel, while none of the other parameters
are guaranteed to be negative (for example $J_{tt}$ is always
positive). Thus for the purpose of mean field analysis, we will
just keep the first term of Eq.~\ref{H1}, and ignore all the rest
three terms of Eq.~\ref{H1}. The mean field Hamiltonian reads
\beqn H_{\MF} &=& \sum_{ij} J_{ts} \left( \vec{\phi} \cdot
\vec{\Delta}^{ts}_{ij} + H.c. \right) - J_{ts} |\vec{\phi}|^2
\cr\cr & +& \sum_j \mu (\sum_{\alpha = 1}^4 b^\dagger_{j,\alpha}
b_{j,\alpha} - \kappa), \label{MF}\eeqn where $\vec{\phi}$ is a
complex vector under $\SU(2)^s$. Here we choose a uniform ansatz
of $\vec{\phi}$ on the entire lattice, and all the links $i,j$ are
included in the sum with the convention $j = i + \hat{e}$ with
$\hat{e} = (1,0)$, $(-1/2, \pm \sqrt{3}/2)$, $i.e.$ the mean field
ansatz explicitly preserves the translation and rotation by
$2\pi/3$ symmetry of the triangular lattice, while all the crystal
symmetries are preserved as projected symmetry group (PSG). $\mu$
is another variational parameter of the mean field calculation
which guarantees that the filling of Schwinger boson is fixed at
$\kappa$ on every site.

The $\FM \otimes 120^\circ$ ordered state corresponds to the mean
field ansatz with $\vec{\phi} = \vec{\phi}_1 + \ii \vec{\phi}_2$,
and the real vectors $\vec{\phi}_1$ and $\vec{\phi}_2$ {\it
orthogonal} with each other. For example, when $\vec{\phi} = \phi
(1, \ii, 0)$, only the spin-up ($\sigma^3 = +1$) Schwinger bosons
participate in this mean field analysis. The ferromagnetic order
parameter corresponds to the following gauge invariant
quantity:\beqn \vec{M} \sim \ii \vec{\phi} \times \vec{\phi}^\ast
\sim \vec{\phi}_1 \times \vec{\phi}_2. \eeqn

With only spin-up Schwinger bosons, the mean field calculation
reduces precisely to the SU(2) spin-1/2 Heisenberg model on the
triangular lattice~\cite{SLsachdev} with Heisenberg coupling
$J_{ij} = - 4J_{ts}$ (The $\Sp(N)$ Heisenberg model defined in
Ref.~\onlinecite{SLsachdev} has Hamiltonian $H = \sum_{i,j} -
\frac{1}{2N} J_{ij} \Delta^\dagger_{ij}\Delta_{ij}$ where
$\Delta_{ij}$ is the $\Sp(N)$ singlet pairing between Schwinger
bosons on sites $i,j$): \beqn H_{\MF} &=& \sum_{ij} J_{ts} \left(
2 \phi \ b^t_{\uparrow, i} \ii \tau^2 b_{j,\uparrow} + H.c.
\right) - 2 J_{ts} \phi^2 \cr\cr & +& \sum_j \mu (\sum_{\alpha =
1}^2 b^\dagger_{j,\uparrow, \alpha} b_{j,\uparrow, \alpha} -
\kappa). \label{MF2}\eeqn Because the spin-down Schwinger bosons
do not contribute to the mean field decomposition when $\vec{\phi}
\sim (1, \ii, 0)$, we replace the constraint in Eq.~\ref{cons} by
$\sum_{\alpha = 1}^2 b^\dagger_{j,\uparrow, \alpha} b_{j,\uparrow,
\alpha} = \kappa$ in Eq.~\ref{MF2}. Now technically the mean field
theory Eq.~\ref{MF2} corresponds to the ``zero-flux state" in
Ref.~\onlinecite{SLwang}, which has lower mean field energy than
other mean field ansatz~\cite{SLwang} for this nearest neighbor
model, and it makes the minima of the Schwinger boson band
structure locate at the corner of the Brillouin zone $\vec{Q} =
(\pm 4\pi/3, 0)$. The mean field solution gives $\mu > 0$, which
is consistent with the fact that we set $\sum_{\alpha=1}^2
b^\dagger_{j,\downarrow,\alpha}b_{j,\downarrow,\alpha} = 0$. And
at the mean field level, when the filling of the Schwinger boson
$\kappa$ is greater than 0.34~\cite{SLsachdev}, $b_\alpha$
condenses, which leads to a fully polarized FM in the spin space,
and also $120^\circ$ state in the valley space.

If the mean field value of $\vec{\phi}$ is real (or equivalently
if $\vec{\phi}_1$ is parallel to $\vec{\phi}_2$, for example,
$\vec{\phi} \sim \phi (0, 0, 1)$, both spin-up and spin-down
Schwinger bosons participate in the mean field analysis, and the
mean field analysis is technically equivalent to the calculations
in Ref.~\onlinecite{SLsachdev} for the $\Sp(2) \sim \SO(5)$
antiferromagnet on the triangular lattice also with Heisenberg
coupling $J_{ij} = - 4J_{ts}$, because $\left(
\vec{\Delta}^{ts}_{ij}, \vec{\Delta}^{st}_{ij} \right)$ together
form a $\SO(6)$ vector, and condensing each component of the
vector breaks the $\SO(6)$ down to $\SO(5) \sim \Sp(2)$. Each
component of $\vec{\Delta}_{ts}$ can be viewed as the $\Sp(2)$
singlet introduced in Ref.~\onlinecite{SLsachdev}: \beqn H_{\MF}
&=& \sum_{ij} J_{ts} \left( \phi \ b^t_{i} \ii \sigma^{12} b_{j} +
H.c. \right) - J_{ts} \phi^2 \cr\cr & +& \sum_j \mu (\sum_{\alpha
= 1}^4 b^\dagger_{j, \alpha} b_{j,\alpha} - \kappa) ,
\label{MF3}\eeqn Quoting the results in
Ref.~\onlinecite{SLsachdev}, the $\FM \otimes 120^\circ$ state
with the previous mean field ansatz with $\phi_1 \perp \phi_2$ has
a lower mean field ground state energy density, which is
consistent with our analytical observation and also numerical
simulation.

{\it ---Zeeman field driven Metal-Insulator transition}

Since the insulator has a fully polarized ferromagnetic order, its
energy can be tuned by an external Zeeman field. An inplane
magnetic field, whose main effect is the Zeeman coupling can drive
a first order metal-insulator transition (a level-crossing)
between the unpolarized metal and the fully polarized
ferromagnetic insulator, as was observed experimentally at the 1/4
and 3/4 filling of TDBG~\cite{kimtalk,TDBG2}.

There is another possible mechanism of metal-insulator transition
driven by a Zeeman field. At the metalic side at the transition,
the system is likely described by a $t-J$ model with a similar
$J,J^s,J^v$ terms as Eq.~\ref{H}. The Zeeman field tends to
polarize the spin, which effectively increases the
antiferromagnetic coupling in the valley space $J^v_{eff} = J^v +
J \langle \vec{\sigma}_i \cdot \vec{\sigma}_j \rangle$. Thus at
certain temperature, the magnitude of the $120^\circ$ order in the
valley space is tunable and enhanced by an external Zeeman field.
If the insulating behavior of the system is a consequence of the
finite momentum valley order which folds the Brillouin zone and
partially gaps out the Fermi surface, an increasing magnitude of
the $120^\circ$ order in the valley space can gap out larger
portion of the Fermi surface, decrease the charge carrier density,
and hence eventually drive a {\it continuous} metal insulator
transition.

\section{Liquids and topological phases}

When $J^v \sim J^s \sim J$, it would be rather difficult for the
system to form any semiclassical order due to ``double
frustration": the $J^s$ and $J^v$ term of Eq.~\ref{H} are both
already frustrated due to the geometry of the triangular lattice,
while the $J$ term further frustrates/disfavors the simultaneous
$120^\circ$ semiclassical order of $\vec{\sigma}_i$ and
$\vec{\tau}_i$. Since there is an obvious Lieb-Shultz-Matthis
theorem which forbids a completely trivial disordered phase, we
expect this ``double frustration" effect to lead to either a
completely disordered spin-valley liquid state, or a partially
ordered state with certain topological order. In this section we
explore several possible spin-valley liquids or topological orders
in the region $J^v \sim J^s \sim J$.

{\it ---Spin nematic $Z_2$ topological phase}

Let us get back to the mean field Hamiltonian Eq.~\ref{MF}. As we
discussed before, if the mean field value of $\vec{\phi}$ is real
(or equivalently if $\vec{\phi}_1$ is parallel to $\vec{\phi}_2$,
for example, $\vec{\phi} \sim (0, 0, 1)$, both spin-up and
spin-down Schwinger bosons participate in the mean field analysis,
and the mean field analysis is technically equivalent to the
calculations in Ref.~\onlinecite{SLsachdev} for the $\Sp(4)$
antiferromagnet on the triangular lattice. And with large spin
symmetry, the quantum fluctuation makes it more difficult for
$b_\alpha$ to condense. If $b_\alpha$ is not condensed, the mean
field order parameter $\vec{\phi}$ already breaks the $\SU(2)^s$,
and also break the $\U(1)$ gauge symmetry down to $Z_2$ gauge
degree of freedom.

The nature of the state with condensed $\vec{\phi}$ but
uncondensed $b_\alpha$ depends on the nature of $\vec{\phi}$ under
time-reversal. The transformation of $b_\alpha$ under
time-reversal can be inferred by the fact that $\vec{\sigma}
\rightarrow - \vec{\sigma}$, $(\tau^1, \tau^2, \tau^3) \rightarrow
(\tau^1, \tau^2, - \tau^3)$: \beqn \mathcal{T}: b_j \rightarrow
\ii \sigma^{21} b_j,  \ \ \ \vec{\Delta}^{ts}_{ij} \rightarrow
\vec{\Delta}^{ts}_{ij}, \eeqn as long as $\vec{\phi}$ is a real
vector (or $\vec{\phi}_1$ parallel with $\vec{\phi}_2$),
time-reversal is preserved, and this state is a spin nematic $Z_2$
topological order. By contrast, if $\vec{\phi}_1 \perp
\vec{\phi}_2$ the time-reversal is broken.

{\it ---$Z_2 \times Z_2$ spin-valley liquid}

More states can be constructed by introducing two flavors of
Schwinger bosons $b^s_{j,\alpha}$ and $b^v_{j,\alpha}$ for the
spin and valley space on each site respectively, which are subject
to the constraint \beqn \sum_{\alpha = 1,2}
b^{s,\dagger}_{j,\alpha}b^{s}_{j,\alpha} =
b^{v,\dagger}_{j,\alpha} b^{v}_{j,\alpha} = 1. \label{sb} \eeqn
These two constraints introduces two $\U(1)$ gauge symmetries. It
is fairly straightforward to construct the $\FM \otimes 120^\circ$
state using this type of Schwinger bosons: $b^{s}_\alpha$
condenses at zero momentum, while simultaneously $b^v_\alpha$
condenses at the corner of the Brillouin zone.

In fact, due to the ``double frustration" effect, both the spin
and valley space can form a $Z_2$ topological order (overall
speaking the system is in a $Z_2\times Z_2$ spin-valley liquid
state), whose $e$ particles carry the fundamental representation
of $\SU(2)^s$ and $\SU(2)^v$ respectively, as long as neither
$b^{s}_{j,\alpha}$ nor $b^{v}_{j,\alpha}$ introduced in
Eq.~\ref{sb} condenses when the mean field parameters break both
$\U(1)$ gauge symmetries down to $Z_2$.


Starting from the $Z_2 \times Z_2$ spin-valley liquid state, one
can also construct a spin-valley liquid with only one $Z_2$
topological order. This can be formally obtained by forming bound
state of the ``visons" (the $m$ excitations) of both $Z_2$
topological orders, and condense the bound state. This condensate
will confine $b^s_{\alpha}$ and $b^v_{\alpha}$ separately, but
their bound state is still deconfined, and becomes the $e$
particle of the new $Z_2$ topological order. This final $Z_2$
topological order preserves all the symmetries of the system, and
it can also be constructed using the same mean field formalism as
Eq.~\ref{H1}, as long as one condenses the
spin-singlet/valley-singlet pairing operator $\Delta^{ss}_{ij}$ in
Eq.~\ref{H1}.

{\it ---$\U(1) \times \U(1)$ Dirac spin-valley liquid}

More exotic spin-valley liquid states can be constructed by
introducing fermionic slave particles $f^s_{j,\alpha}$ and
$f^v_{j,\alpha}$ which are subject to the constraints \beqn
\sum_{\alpha = 1,2} f^{s,\dagger}_{j,\alpha}f^{s}_{j,\alpha} =
f^{v,\dagger}_{j,\alpha} f^{v}_{j,\alpha} = 1. \eeqn In
Ref.~\onlinecite{luz2}, a Dirac spin liquid with $\U(1)$ gauge
field and $N_f = 4$ flavors of Dirac fermions was constructed for
spin-1/2 systems on the triangular lattice. And this Dirac spin
liquid is the parent state of both the $120^\circ$ ordered state
and the valence bond solid state~\cite{luz2,songmono1,songmono2},
and it could be a deconfined quantum critical point between these
two different ordered states~\cite{xutriangle}.

In our case, both spin and valley space can form the Dirac liquid
phase mentioned above, due to the double frustration effect. Thus
in total there are eight flavors of Dirac fermions and two $\U(1)$
gauge fields.

{\it --- The SU(4) point}

At the point $J^v = J^s = J$, this model has a $\SU(4) \sim
\SO(6)$ symmetry. Although semiclasical approach such as nonlinear
sigma model were studied before for $\SU(N)$ antiferromagnet with
other representations~\cite{readsachdev}, with a fundamental
representation on every site, this model has no obvious
semiclassical limit to start with, and it is expected to be a
nontrivial spin liquid or topological order. At this point, it is
most convenient to define a four component Schwinger boson
$b_{j,\alpha}$ on every site which forms fundamental
representation under both the spin and valley $\SU(2)$ symmetry,
and there is a constraint $\sum_{\alpha = 1}^4
b^\dagger_{j,\alpha}b_{j,\alpha} = 1$.

Unlike a $\SU(2)$ spin system, one can prove that at the $\SU(4)$
point there cannot be a fully symmetric $Z_2$ spin liquid whose
$e$ particle is the $b_\alpha$ slave particle. The reason is that
all the local spin excitations can be written as
$b^\dagger_{j,\alpha} b_{j,\beta}$ with different $\alpha,\beta =
1 \cdots 4$, hence all the local spin excitations are invariant
under the $Z_4$ center of the $\SU(4)$ group. In a $Z_2$
topological order, two of the $e$ particles should merge into a
local excitations, while two $b_\alpha$ slave particle cannot fuse
into a representation that is invariant under the $Z_4$ center.
This argument also shows that a $Z_2$ topological order whose $e$
particle is a $\SO(6)$ vector is allowed.

On the other hand, using the slave particle $b_\alpha$ one can
construct a $Z_2$ topological order with certain spontaneous
$\SU(4)$ symmetry breaking. At the $\SU(4)$ point, the model
Eq.~\ref{H} can be written as \beqn H = \sum_{ij} J \left( -
\frac{5}{4} (\vec{\Delta}_{ij}^\dagger) \cdot (\vec{\Delta}_{ij})
+ \cdots \right), \eeqn where $\vec{\Delta}_{ij}$ is a six
component vector pairing between $b_{\alpha}$. One can introduce a
six component complex $\SO(6)$ vector mean field parameter
$\vec{\phi}$: \beqn H_{\MF} = \sum_{ij} J \left( - \frac{5}{4}
\vec{\phi}\cdot \vec{\Delta}_{ij} + H.c. \right) + \frac{5}{4} J
|\vec{\phi}|^2. \eeqn The complex vector $\vec{\phi} =
\vec{\phi}_1 + \ii \vec{\phi}_2$, where its real and imaginary
parts $\vec{\phi}_1$ and $\vec{\phi}_2$ can be either parallel or
orthogonal to each other. If the Schwinger boson does not
condense, both mean field theories would lead to a $Z_2$
topological order on top of the spontaneous $\SU(4)$ symmetry
breaking.


Our DMRG simulation actually suggest that the $\SU(4)$ point of
the spin-valley model is a spin-valley liquid state with a Fermi
surface of fermionic slave particles, which will be presented in
detail in another work.

\section{Conclusion}

In this work we demonstrated both analytically and numerically
that a quantum spin-valley model with all antiferromagnetic
interaction can have a fully polarized ferromagnetic order in its
phase diagram. We propose possible mechanism for an inplane Zeeman
field to drive a metal-insulator transition, as was observed
experimentally at the 1/4 and 3/4 filling of TDBG. We also
discussed various possible nontrivial spin-valley liquid state and
topological order of this model.

We would like to acknowledge several previous theoretical works
that studied the ferromagnetism in Moir\'{e} systems using
different approaches and different
models~\cite{vafek2,bruno,zaletel,TDBGt2}. For example, in
Ref.~\onlinecite{bruno}, a spin-valley model with {\it
ferromagnetic couplings} on an effective honeycomb Moir\'{e}
lattice was derived for the twisted bilayer graphene system. While
our work (which aims to understand a different Moir\'{e} system,
$i.e.$ the twisted double bilayer graphene) demonstrated that
ferromagnetism can emerge from the spin-valley model on a
triangular lattice with {\it fully antiferromagnetic interaction}.

Within our framework, under doping, again the system is likely
described by the $t-J$ model with the similar $J,J^s,J^v$ terms as
Eq.~\ref{H}. Then the analysis in Ref.~\onlinecite{xuleon} still
applies: the spin-triplet/valley-singlet pairing channel between
electrons would become the strongest pairing channel. Due to a
strong on-site Hubbard interaction, the system would still prefer
to become a $d + id$ topological superconductor with spin triplet
pairing.

Anna Keselman and Chao-Ming Jian are supported by the Gordon and
Betty Moore Foundations EPiQS Initiative through Grant GBMF4304.
Cenke Xu is supported by the David and Lucile Packard Foundation.
Use was made of the computational facilities administered by the
Center for Scientific Computing at the CNSI and MRL (an NSF MRSEC;
DMR-1720256) and purchased through NSF CNS-1725797.

\bibliography{FM}

\end{document}